\documentclass[a4paper,11pt]{article}
\usepackage[position=t,singlelinecheck=off]{subfig}
\usepackage{amssymb,amsmath,graphics,graphicx,float,braket}
\usepackage{array,multirow,setspace,makecell,pbox}
\def\beq{\begin{equation}}
\def\eeq{\end{equation}}
\def\bea{\begin{eqnarray}}
\def\eea{\end{eqnarray}}
\def\nn{\nonumber}

\makeatletter
\newcommand{\lvast}{\bBigg@{3}}
\newcommand{\svast}{\bBigg@{2}}
\newcommand{\vast}{\bBigg@{4}}
\newcommand{\Vast}{\bBigg@{5}}
\makeatother
\makeatletter
\def\@cite#1#2{${\mbox{#1\if@tempswa , #2\fi}}$}
\makeatother

\newcolumntype{P}[1]{>{\centering\arraybackslash}p{#1}}
\newcolumntype{M}[1]{>{\centering\arraybackslash}m{#1}}
\begin{document}
\thispagestyle{empty}
\begin{center}
\begin{LARGE}
\textsf{Evolution of the entanglement of the} $N00N$\textsf{-type of states in a coupled
two cavity system via an adiabatic approximation}
\end{LARGE} \\

\bigskip\bigskip
R. Chakrabarti${}^1$\footnote{E-mail: ranabir@imsc.res.in},
G. Sreekumari${}^{2,3}$
\footnote{E-mail: gsree64@gmail.com}  
and V. Yogesh${}^{2}$
\footnote{E-mail: yo.physics@gmail.com}  
\\
\begin{small}
	\bigskip
	\textit{
		${}^1$The Institute of Mathematical Sciences, CIT Campus, Taramani, Chennai 600 113, India.\\
		${}^{2}$Department of Theoretical Physics, 
		University of Madras, Maraimalai Campus,\\ Guindy,
		Chennai 600 025, India. \\
		${}^{3}$Department of Physics, Loyola College, Chennai 600 034, India.}\\
		
\end{small}
\end{center}

\vfill
\begin{abstract}
We study a system of two cavities each encapsulating a qubit and an oscillator degrees of freedom. An ultrastrong interaction strength between the qubit and the oscillator is assumed, and the photons are allowed to hop between the cavities. A partition of the time scale between the fast moving oscillator and the slow moving qubit allows us to set up an adiabatic approximation procedure where we employ the delocalized degrees of freedom to diagonalize the Hamiltonian. The time evolution of the $N00N$-type initial states now furnishes, for instance, the reduced density matrix of a bipartite system of two qubits. For a macroscopic size of the $N00N$ component of the initial state the sudden death of the entanglement between the qubits and its continued null value  are  prominently manifest as the information percolates to the qubits after long intervals. For the low photon numbers of the initial states the dynamics produces almost maximally entangled two-qubit states, which by  utilizing the Hilbert-Schmidt 
distance between the density matrices, are observed to be nearly pure generalized Bell states. 
\end{abstract}
\section{Introduction}
\label{Intro}
The physical structure of the cavity and circuit quantum electrodynamics is represented by  the localized oscillator modes interacting with the two level systems. Models involving coupled arrays of the qubit-oscillator degrees of freedom, where the photons are permitted to hop between the cavities, recently attracted much experimental and theoretical attention. Various experimental advances in areas such as the photonic crystals [\cite{AKSV2003}], the optical microcavities containing the highly localized defect modes within the
photonic band gap [\cite{BTO2000}], and superconducting devices [\cite{ABS2002}, \cite{WALLRAFF2004}] triggered many studies of these arrays. Such formations have been recently considered for providing a framework for the distributed quantum computation [\cite{CEHM1999}], the generation of entanglement [\cite{AB2007}], the transport of a quantum state [\cite{BAB2007}-\cite{CS2011}], and the cluster state quantum computation that uses the polaritonic excitations[\cite{HBP2007}, \cite{AK2008}]. 
 
\par

 The qubit-oscillator interaction has been studied extensively under the Jaynes-Cummings model [\cite{JC1963}] that employs  the rotating wave approximation holding good for the regime characterized by a weak coupling as well as a small detuning between the qubit and the oscillator frequencies. Recent experiments, however, probe the ultrastrong  coupling domain, where the rotating wave approximation is not valid.   Experimental realizations such as a  metal-dielectric-metal microcavity combined with quantum well intersubband transitions   generating the cavity polariton states in the terahertz region [\cite{TODOROV2009}, \cite{HIF2012}],  a quantum semiconductor microcavity displaying specific signatures of the ultrastrong coupling regime of the light-matter interaction [\cite{GUNTER2009}, \cite{AAA2009}], a nanoelectromechanical resonator capacitively coupled to a Cooper-pair box driven by the microwave currents [\cite{ABS2002}, \cite{WALLRAFF2004}, \cite{LSESR2009}], a flux-biased quantum circuit that utilizes the large inductance of a Josephson junction to produce an ultrastrong coupling with a coplanar waveguide resonator [\cite{HOFHEINZ2009}, \cite{NIEMCZYK2010}] fall in this group. In particular, the superconducting qubits and circuits facilitate wide range of variability of the parameters, and, consequently, may be chosen as the preferred  building blocks for the quantum simulators [\cite{YN2005}-\cite{GAN2014}].  Moreover, the integrated hybrid quantum circuits involving the atoms, spins, cavity photons and the superconducting qubits with the nanomechanical resonators may significantly contribute towards the fabrication of interfaces [\cite{ZAYN2013}] in the quantum communication network.
 
\par
 
The Hamiltonian of the strongly coupled qubit-oscillator system embodies terms that do not preserve the total excitation number. To analyze them in the regime where the high oscillator frequency dominates over the low  qubit frequency, the authors of [\cite{IGMS2005}, \cite{AN2010}] have advanced an adiabatic approximation scheme that exploits the separation of the slow and the fast changing degrees of freedom. This validates the decoupling of the full bipartite Hamiltonian into components related to each time scale, and permits its approximate diagonalization [\cite{IGMS2005}]. Utilizing the adiabatic approximation the energy eigenvalues and the eigenstates of the physical systems comprising of two [\cite{YZZ2012}, \cite{D2016}] and three [\cite{SCWY2014}] qubits coupled with a \textit {single} oscillator degree of freedom have been studied. 

\par

In another development, much notice is devoted [\cite{BIWH1996},\cite{KLD2002}] to a bipartite, path-entangled,  Schr\"{o}dinger cat type discrete photon number state, commonly called the $N00N$ state, where  a fixed finite number of photons  are all in either of the two available modes.  These states endowed with the multiphoton excitations possess the same degree of entanglement as the Bell states. High precision phase measurement may be accomplished by harnessing the multiphoton entangled $N00N$ states, where a higher photon number leads to increased advantage. In particular, these states facilitate [\cite{BIWH1996}-\cite{MLS2004}] achieving the optimal accuracy permitted by the Heisenberg uncertainty principle. The enhanced phase sensitivity of these states is employed towards reaching the sub-Rayleigh resolution in quantum lithography [\cite{BOTO2000}]. Additionally, utilizing  the $N00N$ component of an entangled four-photon state precise optical phase measurement with a visibility that surpasses the accuracy limit obtainable with the unentangled photons has  been experimentally realized [\cite{NOOST2007}]. The optical $N00N$ states with high photon numbers, which, therefore, tend towards the macroscopic entangled states, have recently been generated [\cite{AAS2011}] using the multiphoton interference of quantum down-converted light with a classical coherent state. Employing a superconducting quantum circuit that includes the Josephson qubits coupled with the two independent  microwave resonators the entangled multiphoton $N00N$ states have also been achieved [\cite{Wang2011}].
 
\par

In the overall context it is important to study the evolution of various nonclassical states in a cluster of coupled cavity systems. Recently the authors of Ref. [\cite{OIK2008}] considered the dynamics of a two-site coupled cavity model over a large range of values of the qubit-cavity detuning and the photon tunneling strength. Describing the electromagnetic fields and the spin operators in the system via the delocalized modes they investigated the atomic state transfer. The qubit-oscillator interaction, however,  is characterized [\cite{OIK2008}] by the rotating wave approximation [\cite{JC1963}]  that  preserves the total number of excitations. Making a departure, we, in this work admit strong coupling of the qubit-oscillator hybrid system
where the conservation of the total number of excitations is not assumed. For specificity, we consider the evolution of a 
$N00N$-type of state in a structure comprising  of two coupled qubit-oscillator systems. In Sec. \ref{Eigenstate} we enlist the delocalized coordinates in conjunction with the adiabatic approximation procedure to diagonalize the Hamiltonian. The evolution of the $N00N$-type of state described in Sec. \ref{N00N_evolution} allows us to construct (Sec. \ref{reduced}) the bipartite reduced density operator for the qubits. The time-variation of the entanglement of the qubits is studied utilizing the concurrence [\cite{W1998}] as the measure. We conclude in Sec. \ref{conclu}.

\section{Diagonalizing the Hamiltonian  via the adiabatic approximation}
\label{Eigenstate}
\setcounter{equation}{0}
We consider two identical cavities each containing a  two-level atom that is strongly coupled to a localized  oscillator degree of freedom, where the Hamiltonian in natural units $(\hbar = 1)$ reads 
\beq
H=\sum_{\jmath=0,1} 
\left( -\dfrac{\Delta}{2} \sigma_{\jmath}^{\mathrm{x}} + \omega \,  a_{\jmath}^{\dagger}  a_{\jmath}
+  \lambda \, \sigma_{\jmath}^{\mathrm{z}} \left( a_{\jmath} + a_{\jmath}^{\dagger} \right)  \right) +
\nu \left( a_{0}^{\dagger} a_{1} + a_{0}  a_{1}^{\dagger}   \right).
\label{hamiltonian}
\eeq
 The harmonic oscillator modes $\{ a_{\jmath}, a_{\jmath}^{\dagger}, \hat{n}_{j} \equiv a_{\jmath}^{\dagger}  a_{\jmath} |\, \jmath \in (0,1)\}$ are characterized by the frequency 
 $\omega$, and the  qubit variables described by the Pauli spin operators $\{\sigma_{\jmath}^{\mathcal{X}}| \,\jmath \in (0,1), \mathcal{X} = \mathrm{x,y,z}\}$ possess the energy splitting parameter $\Delta$. The qubit-oscillator coupling strength is denoted by $\lambda$, whereas the two cavities are interlinked via the photon hopping parameter $\nu$. To facilitate our analysis we now recast the Hamiltonian (\ref{hamiltonian}) using the delocalized field and atomic modes, which are given by the symmetric and the antisymmetric  linear combinations of  their local analogs pertaining to a cavity: 
\bea
A_{0} = \dfrac{1}{\sqrt{2}} \left( a_{0} + a_{1} \right),  A_{1} = \dfrac{1}{\sqrt{2}} \left( a_{0} - a_{1} \right), 
S_{0}^{\mathcal{X}} = \dfrac{1}{\sqrt{2}} \left( \sigma_{0}^{\mathcal{X}} + \sigma_{1}^{\mathcal{X}} \right), 
S_{1}^{\mathcal{X}} = \dfrac{1}{\sqrt{2}} \left( \sigma_{0}^{\mathcal{X}} - \sigma_{1}^{\mathcal{X}} \right), 
\mathcal{X} = \mathrm{x,y,z}.
\label{delocal_var}
\eea
 The delocalized oscillator modes obeying the commutation relation $\big\{\big[ A_{\jmath}, A_{\ell}^{\dagger} \big] = \delta_{\jmath \ell}; \; \jmath,\ell =0,1\big\}$, and the corresponding spin variables introduced above transform the  Hamiltonian (\ref{hamiltonian}) as follows:
\beq
H=H_{\mathcal{Q}}  + \Omega_{0} A_{0}^{\dagger} A_{0} +  \Omega_{1} A_{1}^{\dagger} A_{1} 
+ \lambda \left( S_{0}^{\mathrm{z}} \left( A_{0} + A_{0}^{\dagger}\right) 
+ S_{1}^{\mathrm{z}} \left( A_{1} + A_{1}^{\dagger}\right) \right), \; 
H_{\mathcal{Q}} = -\dfrac{\Delta}{\sqrt{2}} S_{0}^{\mathrm{x}},
\label{H_delocal}
\eeq
where the tunneling of the photons between the cavities lifts the degeneracy of the frequencies of the delocalized quanta:
$ \Omega_{0} = \omega + \nu, \Omega_{1} = \omega - \nu$.
It has been noted [\cite{IGMS2005}] that the parity operator conserves the qubit-oscillator Hamiltonian. For the two-cavity example (\ref{hamiltonian}) studied here the parity operator assumes the form expressed via the localized and the delocalized variables, respectively, as  
\beq
P = \exp \left( i \pi (a_{0}^{\dagger} a_{0} + a_{1}^{\dagger} a_{1}) + i\frac{\pi}{2} 
(\sigma_{0}^{\mathrm{x}} + \sigma_{1}^{\mathrm{x}} ) \right) \Longrightarrow
P = \exp \left( i \pi (A_{0}^{\dagger} A_{0} + A_{1}^{\dagger} A_{1}) + i\frac{\pi}{\sqrt{2}} 
S_{0}^{\mathrm{x}}  \right),
\label{P_delocal}
\eeq
where the commutation property $[P, H]= 0$ is preserved.

\par

To implement our construction of the evolution of the $N00N$-type states, we, following [\cite{IGMS2005}, \cite{AN2010}], now proceed towards  the diagonalization process of the Hamiltonian (\ref{H_delocal}) in the adiabatic approximation scheme that  has been found to be appropriate in the large detuning limit $(\Delta \ll \omega)$ as it utilizes the difference between the time scales of the slow-moving atomic modes and that of the fast-moving oscillators. The high-frequency oscillators are assumed to instantaneously adjust to the slow-changing state of the qubit observables 
$\{\sigma_{\jmath}^{\mathrm{z}}| \,\jmath \in (0,1)\}$ so that the construction permits, in the course of diagonalization of the oscillator modes, replacing the spin-variables with the corresponding eigenvalues: $\{\braket{\sigma_{\jmath}^{\mathrm{z}}} = m_{\jmath} = \pm 1| \,\jmath \in (0,1)\}$. The delocalized spin variables introduced in (\ref{delocal_var}) now admit the 
substitution
\beq
\braket{S_{0}^{\mathrm{z}}}  =  \dfrac{1}{\sqrt{2}} 
\braket{\left( \sigma_{0}^{\mathrm{z}} + \sigma_{1}^{\mathrm{z}}  \right)} 
=  \dfrac{1}{\sqrt{2}} \left( m_{0} + m_{1} \right), \quad
\braket{S_{1}^{\mathrm{z}}}  =  \dfrac{1}{\sqrt{2}} 
\braket{\left( \sigma_{0}^{\mathrm{z}} - \sigma_{1}^{\mathrm{z}}  \right)} 
=  \dfrac{1}{\sqrt{2}} \left( m_{0} - m_{1} \right)
\label{delocal_expect}
\eeq
that expresses the effective Hamiltonian $H_{\mathcal{O}}$ of the oscillator degrees of freedom in the following form:
\beq
 H_{\mathcal{O}} =  \Omega_{0} \left( A_{0}^{\dagger} A_{0} 
 + \mu_{0} \left( A_{0} + A_{0}^{\dagger} \right) \right) +  \Omega_{1} \left( A_{1}^{\dagger} A_{1} 
 + \mu_{1} \left( A_{1} + A_{1}^{\dagger} \right) \right),
\label{H_oscillator} 
\eeq
where the coefficients read
$\mu_{0} = \frac{\lambda}{\sqrt{2} \Omega_{0}} (m_{0} + m_{1}), \; 
\mu_{1} = \frac{\lambda}{\sqrt{2} \Omega_{1}} (m_{0} - m_{1})$.
The displacement operators $D_{0}(\mu_{0}) = \exp \big( \mu_{0} \big( A_{0}^{\dagger} - A_{0} \big) \big), \,
D_{1}(\mu_{1}) = \exp \big( \mu_{1} \big( A_{1}^{\dagger} - A_{1} \big) \big)$ acting on the phase space of the oscillator variables now facilitates the recasting of the effective Hamiltonian (\ref{H_oscillator}) as 
\beq
H_{\mathcal{O}} = \Omega_{0} D_{0}(\mu_{0})^{\dagger} A_{0}^{\dagger} A_{0} D_{0}(\mu_{0})
+ \Omega_{1} D_{1}(\mu_{1})^{\dagger} A_{1}^{\dagger} A_{1} D_{1}(\mu_{1}) - \Omega_{0} \mu_{0}^{2} 
-\Omega_{1} \mu_{1}^{2}.
\label{H_diplace}
\eeq
The eigenstates of the oscillator component (\ref{H_diplace}) of the Hamiltonian now readily follows as the displaced number states corresponding to the delocalized degrees of freedom: $\hat{N}_{0} \equiv  A_{0}^{\dagger} A_{0}, 
\hat{N}_{1} \equiv  A_{1}^{\dagger} A_{1}, \hat{N}_{0} \ket{N_{0}} = N_{0} \ket{N_{0}}, 
\hat{N}_{1} \ket{N_{1}} = N_{1} \ket{N_{1}}$. The eigenstates of the Hamiltonian (\ref{H_diplace}) explicitly read 
\beq
D_{0}(\mu_{0})^{\dagger} D_{1}(\mu_{1})^{\dagger} \ket{N_{0},N_{1}} = \ket{N_{0,m_{0}+m_{1}},N_{1,m_{0}-m_{1}}},\quad
\ket{N_{0}}= \frac{(A_{0}^{\dagger})^{N_{0}}}{\sqrt{N_{0}!}} \ket{0},\;
\ket{N_{1}}= \frac{(A_{1}^{\dagger})^{N_{1}}}{\sqrt{N_{1}!}} \ket{0}.
\label{osc_eigenstate}
\eeq

\par

After completing the approximate diagonalization  the high frequency oscillator components of the Hamiltonian (\ref{H_delocal}),  we now attend to the corresponding low frequency qubit parts. A tensor product of  the qubit states with the displaced oscillator basis states  
\beq
\ket{\Psi_{\braket{m}}^{\braket{N}}} =\ket{N_{0,m_{0}+m_{1}},N_{1,m_{0}-m_{1}};m_{0},m_{1}}, \quad \braket{N} \equiv (N_{0}, N_{1}), \; \braket{m} \equiv (m_{0},m_{1})
\label{tensor_state}
\eeq
provides the construction of the relevant matrix elements of the Hamiltonian (\ref{H_delocal}). For a dominant oscillator frequency $\Delta \ll \omega$ one may neglect [\cite{IGMS2005}] the matrix elements that mix the oscillator states with
different eigenvalues $(N_{0}, N_{1})$ of its number operators. In other words, the separation of oscillator energy levels is much larger than that of the two-level system, and, consequently,  transitions in the two-level system can never trigger an
 excitation of the oscillator. This  approximation truncates the Hamiltonian to a block-diagonal form 
where each block mixes the displaced oscillator states with identical $(N_{0}, N_{1})$ eigenstates of photons. The Hamiltonian for the $(N_{0}, N_{1})$-th block may be represented as follows:
\beq
H^{\braket{N}}=
\begin{pmatrix}
\mathcal{N}-2\dfrac{\lambda ^{2}}{\Omega_{0}} & H_{12}  & H_{13} & 0 \\
H_{21} 
& \mathcal{N}-2\dfrac{\lambda ^{2}}{\Omega_{1}} & 0 & H_{24}  \\
H_{31} & 0 & \mathcal{N}-2\dfrac{\lambda ^{2}}{\Omega_{1}} & H_{34} \\
0 & H_{42} & H_{43} & 
\mathcal{N}-2\dfrac{\lambda ^{2}}{\Omega_{0}}
\end{pmatrix},
\label{H_qubit}
\eeq
where $\mathcal{N} = \Omega_{0} N_{0} + \Omega_{1} N_{1}$. The minor diagonal elements of the block  Hamiltonian 
$H^{\braket{N}}$ vanish as the qubit component $H_{\mathcal{Q}}$ produces only a single spin flip at the order considered here.
The remaining off-diagonal elements are real, and may be evaluated as projections in the Hilbert space:
\bea
H_{12} \!\!\! &=& \!\!\! H_{21} = 
\braket{N_{_{0,-2}},N_{_{1}};-1,-1|H_{_{\mathcal{Q}}}| N_{_{0}},N_{_{1,-2}};-1,1} =  -\dfrac{\Delta}{2} \braket{N_{_{0,-2}}|N_{_{0}}} 
\braket{N_{_{1}}|N_{_{1,-2}}}, \nn \\
H_{13} \!\!\! &=& \!\!\! H_{31} = \braket{N_{_{0,-2}},N_{_{1}};-1,-1|H_{_{\mathcal{Q}}}| N_{_{0}},N_{_{1,2}};1,-1} =  -\dfrac{\Delta}{2} \braket{N_{_{0}}|N_{_{0,-2}}} 
\braket{N_{_{1,2}}|N_{_{1}}}, \nn \\
H_{24} \!\!\! &=& \!\!\! H_{42} =
\braket{N_{_{0}},N_{_{1,-2}};-1,1|H_{_{\mathcal{Q}}}| N_{_{0,2}},N_{_{1}};1,1} =  -\dfrac{\Delta}{2} \braket{N_{_{0}}|N_{_{0,2}}} 
\braket{N_{_{1,-2}}|N_{_{1}}}, \nn \\
H_{34}\!\!\! &=& \!\!\! H_{43} =
\braket{N_{_{0}},N_{_{1,2}};1,-1|H_{_{\mathcal{Q}}}| N_{_{0,2}},N_{_{1}};1,1} =  -\dfrac{\Delta}{2} \braket{N_{_{0}}|N_{_{0,2}}} 
\braket{N_{_{1,2}}|N_{_{1}}}.
\label{H_offdiagonal} 
\eea
The reflection property $\braket{N_{_{1,2}}|N_{_{1}}} = \braket{N_{_{1,-2}}|N_{_{1}}},\, 
\braket{N_{_{0,2}}|N_{_{0}}} = \braket{N_{_{0,-2}}|N_{_{0}}}$ ensures the equality of the  off-diagonal elements in (\ref{H_qubit}):
$ H_{12}=H_{13}=H_{24}=H_{34} \equiv \Lambda _{\braket{N}} = - \frac{\Delta}{2} \exp \left(- \Gamma_{+} \right) 
\mathrm{L}_{N_{0}} \left( \frac{2 \lambda ^{2}}{\Omega_{0}^{2}} \right)
\mathrm{L}_{N_{1}} \left( \frac{2 \lambda ^{2}}{\Omega_{1}^{2}} \right)$, where the parameters read
 $\Gamma_{\pm} = \lambda ^{2} \left( \frac{1}{\Omega_{0}} \pm \frac{1}{\Omega_{1}} \right)$ and the Laguerre polynomial follows the usual expansion: $L_{n}(x) = \sum_{k = 0}^{n} 
\,(-1)^{k}\, \binom{n}{k}\,\frac{x^{k}}{k!}$.

\par

The energy eigenvalues of the block Hamiltonian (\ref{H_qubit}) may now be listed as
\beq
\mathcal{E}_{0}^{\braket{N}} = \mathcal{N} - 2 \dfrac{\lambda ^{2}}{\Omega_{0}}, \;
\mathcal{E}_{1}^{\braket{N}} = \mathcal{N} - 2 \dfrac{\lambda ^{2}}{\Omega_{1}}, \;
\mathcal{E}_{\pm}^{\braket{N}} = \mathcal{N} - \Gamma_{+} \pm \chi_{_{\braket{N}}}, \;
\chi_{_{\braket{N}}} = \sqrt{4 \Lambda ^{2}_{\braket{N}} + \Gamma_{-}^{2}}
\label{eigenenergy}
\eeq
and the corresponding eigenstates assume the form
\bea
\ket{\mathcal{E}_{0}^{\braket{N}}} &=& \dfrac{1}{\sqrt{2}} 
\left( \ket{N_{0,2},N_{1};1,1} - \ket{N_{0,-2},N_{1};-1,-1} \right), \nn\\
\ket{\mathcal{E}_{1}^{\braket{N}}} &=& \dfrac{1}{\sqrt{2}} 
\left( \ket{N_{0},N_{1,2};1,-1} - \ket{N_{0},N_{1,-2};-1,1} \right), \nn\\
\ket{\mathcal{E}_{\pm}^{\braket{N}}} &=& \dfrac{1}{2} 
\Bigg[ \sqrt{\dfrac{\chi_{_{\braket{N}}} \mp \Gamma_{-}}{\chi_{_{\braket{N}}}}} 
\left( \ket{N_{0,2},N_{1};1,1} + \ket{N_{0,-2},N_{1};-1,-1}  \right) \nn\\
&&\pm 
\dfrac{\Lambda _{\braket{N}}}{ \left| \Lambda _{\braket{N}} \right|} 
\sqrt{\dfrac{\chi_{_{\braket{N}}} \pm \Gamma_{-}}{\chi_{_{\braket{N}}}}} 
\left( \ket{N_{0},N_{1,2};1,-1} + \ket{N_{0},N_{1,-2};-1,1} \right) \Bigg], 
\label{N-state}
\eea
The above eigenstates of the block-diagonalized Hamiltonian (\ref{H_qubit}) obey the orthonormality property:
\beq
\braket{\mathcal{E}_{\jmath}^{\braket{N}}|\mathcal{E}_{\ell}^{\braket{N'}}} 
= \delta_{\jmath,\ell} \; \delta_{N_{0},N'_{0}} \; \delta_{N_{1},N'_{1}}, \quad \jmath ,\ell \in \{ 0,1,\pm \}.
\label{orthogonal}
\eeq
Moreover, as the eigenstates  (\ref{N-state}) conform to the following requirement 
\beq
\sum_{N_{0},N_{1}=0}^{\infty} 
\left[ \ket{\mathcal{E}_{0}^{\braket{N}}} \bra{\mathcal{E}_{0}^{\braket{N}}}
 + \ket{\mathcal{E}_{1}^{\braket{N}}}  \bra{\mathcal{E}_{1}^{\braket{N}}} 
 + \ket{\mathcal{E}_{+}^{\braket{N}}}  \bra{\mathcal{E}_{+}^{\braket{N}}} 
 + \ket{\mathcal{E}_{-}^{\braket{N}}}  \bra{\mathcal{E}_{-}^{\braket{N}}} \right] 
 =  \mathrm{I}_{\braket{N}} \; \mathrm{I}_{\braket{m}},
 \label{complete}
\eeq
the set $\big\{\ket{\mathcal{E}_{\jmath}^{\braket{N}}}| \jmath \in \{ 0,1,\pm \},( N_{0},N_{1}) 
\in (0, 1, \ldots \infty)\big\}$
provides a complete basis in the Hilbert space. In (\ref{complete}) the unit operators for the oscillator and the spin basis states, respectively, read
\beq
\sum_{N_{0},N_{1}=0}^{\infty} \ket{N_{0},N_{1}} \bra{N_{0},N_{1}} = \mathrm{I}_{\braket{N}}, \quad
\sum_{m_{0},m_{1}=\pm 1} \ket{m_{0},m_{1}} \bra{m_{0},m_{1}} = \mathrm{I}_{\braket{m}}.
\label{unity}
\eeq

\par

The parity quantum numbers of the energy eigenstates (\ref{N-state}) under the adiabatic approximation are observed as follows. The transformation properties of the basis vectors (\ref{tensor_state}) 
\bea
P \ket{N_{0,\pm 2},N_{1};\pm 1, \pm 1} &=& (-1)^{N_{0}+N_{1}+1} \ket{N_{0,\mp 2},N_{1};\mp 1, \mp 1}, \nn\\
P \ket{N_{0},N_{1,\pm 2};\pm 1, \mp 1} &=& (-1)^{N_{0}+N_{1}+1} \ket{N_{0},N_{1,\mp 2};\mp 1, \pm 1} 
\label{P_N}
\eea
impart the parity eigenvalues to the  energy states (\ref{N-state}). We notice that the states
$\big\{\ket{\mathcal{E}_{\jmath}^{\braket{N}}}|\, \jmath \in (0,1)\big\}$ have opposite parity compared to their partners
$\ket{\mathcal{E}_{\pm}^{\braket{N}}}$ as these two sets comprise of the antisymmetric and symmetric linear combinations of the 
vectors (\ref{tensor_state}), respectively: 
\beq
P \ket{\mathcal{E}_{\jmath}^{\braket{N}}} = (-1)^{N_{0}+N_{1}} \ket{\mathcal{E}_{\jmath}^{\braket{N}}}, \;  \jmath \in (0,1),\qquad
P \ket{\mathcal{E}_{\pm}^{\braket{N}}} = (-1)^{N_{0}+N_{1}+1} \ket{\mathcal{E}_{\pm}^{\braket{N}}}.
\label{P_eigenenergy}
\eeq

\par

\begin{figure}[H]
\hfill \includegraphics[width=12.3cm,height=5.8cm]	{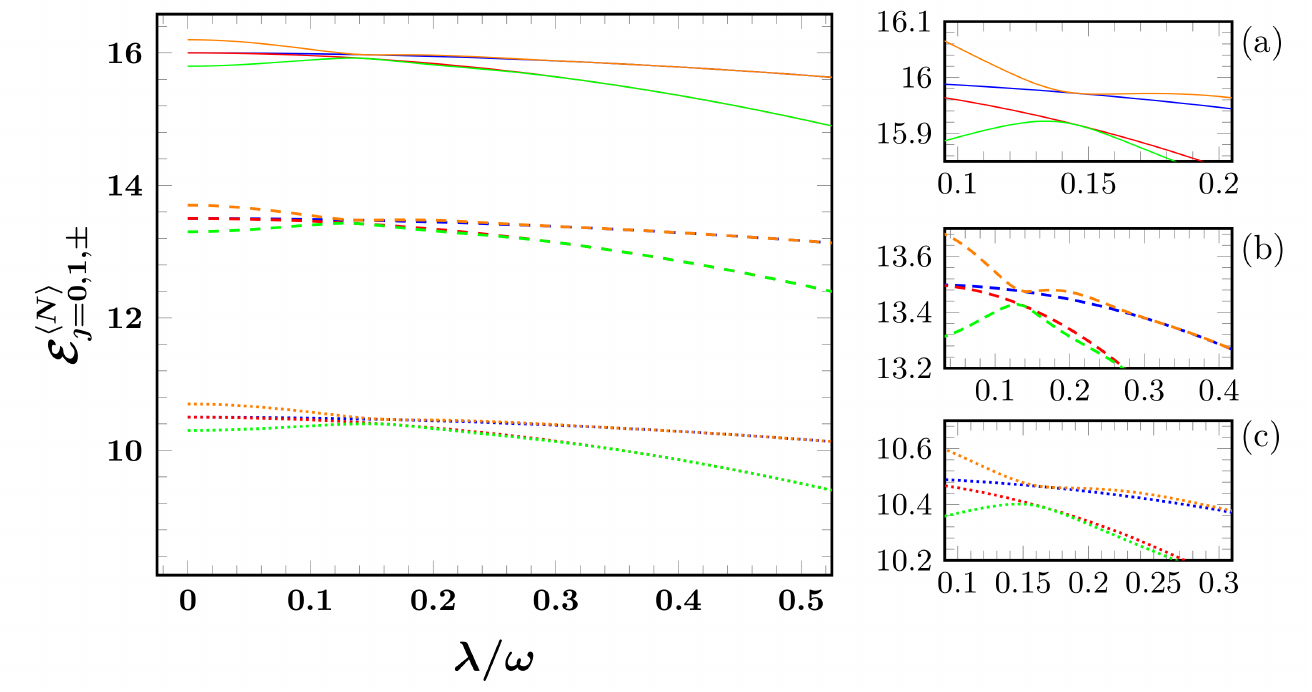} \hspace*{\fill}
\caption{The variations of the energy eigenvalues $\big\{\mathcal{E}_{0}^{\braket{N}} (\mathrm{blue}), 
\mathcal{E}_{1}^{\braket{N}} (\mathrm{red}), \mathcal{E}_{+}^{\braket{N}} (\mathrm{orange}), 
\mathcal{E}_{-}^{\braket{N}}(\mathrm{green})\big\}$ with respect to the coupling strength $\lambda$ are plotted  corresponding to the parametric choices $ \omega=1, \Delta=0.2, \nu=0.5$. The eigenenergies corresponding to the ordered quantum numbers 
 $(N_{0}=5, N_{1}=6), (N_{0}=6, N_{1}=9)$ and $(N_{0}=8, N_{1}=8)$ of the delocalized oscillator  degrees of freedom 
 are represented via   the dotted, dashed and the solid lines, respectively. The coupling strength $\lambda$ regimes where the energy levels become pairwise identical are enlarged on the side of the respective energy bands (Figs. ({\sf a, b, c})).} 
\label{N_level}
\end{figure}

In Fig. \ref{N_level} we plot the energy levels (\ref{eigenenergy}) with varying coupling strength $\lambda$ for different choices of the oscillator quantum numbers $(N_{0}, N_{1})$. As we have retained the photon tunneling
constant to be positive $\nu > 0$, the energy eigenvalues consistent with the parametric range studied here maintain the hierarchy: $\mathcal{E}_{+}^{\braket{N}} \ge \mathcal{E}_{0}^{\braket{N}} \ge \mathcal{E}_{1}^{\braket{N}} \ge 
\mathcal{E}_{-}^{\braket{N}}$. The undulations in the diagrams for $\mathcal{E}_{\pm}^{\braket{N}}$ are manifest due to the presence of the Laguerre polynomials in the corresponding expressions. In the strong coupling  limit $(\lambda \lesssim \omega)$ the energy eigenvalues satisfy  $\mathcal{E}_{+}^{\braket{N}} \rightarrow \mathcal{E}_{0}^{\braket{N}}, \;
\mathcal{E}_{-}^{\braket{N}} \rightarrow \mathcal{E}_{1}^{\braket{N}}$. Corresponding to the zeros of the Laguerre polynomials,
the energies of the opposite parity states $\{\ket{\mathcal{E}_{+}^{\braket{N}}},  \ket{\mathcal{E}_{0}^{\braket{N}}}\}$, 
as well as $\{\ket{\mathcal{E}_{-}^{\braket{N}}},  \ket{\mathcal{E}_{1}^{\braket{N}}}\}$,  become identical. The degeneracy
 of the above two pairs of energy levels are realized in the examples $(N_{0}=5, N_{1}=6), (N_{0}=6, N_{1}=9)$ and 
$(N_{0}=8, N_{1}=8)$ for the coupling strength $\lambda$ equaling $0.166901$, $0.137986$ and $0.145894$, respectively.
They, successively,  correspond to the zeros of the Laguerre polynomials $L_{6}({\mathsf x}), L_{9}({\mathsf x})$  and 
$L_{8}({\mathsf x})$, where ${\mathsf x} = \frac{2 \lambda^{2}}{\Omega_{1}^{2}}$.

\par

\section{Evolution of a localized $N00N$-type of state}
\label{N00N_evolution}
\setcounter{equation}{0}
Towards constructing the time evolution of a $N00N$-type state we first project, in the basis set (\ref{tensor_state}), an arbitrary localized number state $\ket{ n_{0}, m_{0} ;n_{1}, m_{1}}$ obeying the property $\hat{n}_{\jmath} \ket{ n_{0}, m_{0} ;n_{1}, m_{1}} = n_{\jmath}\ket{n_{0}, m_{0} ;n_{1}, m_{1}}, \jmath \in (0, 1)$. These projections read
\bea
\braket{N_{0,\pm 2},N_{1};\pm 1, \pm 1| n_{0}, m_{0} ;n_{1}, m_{1}} & =& (-1)^{n_{0}+n_{1}+N_{1}} \;
\mathcal{C}_{0}^{\langle n;N \rangle}(\pm \lambda) \; \mathcal{F}_{0}^{\langle n;N \rangle} \delta_{m_{0}, \pm 1} \delta_{m_{1}, \pm 1},\nn\\
\braket{N_{0},N_{1,\pm 2};\pm 1, \mp 1| n_{0}, m_{0} ;n_{1}, m_{1}} & =& (-1)^{n_{0}+N_{0}} \;
\mathcal{C}_{1}^{\langle n;N \rangle}(\pm \lambda) \; \mathcal{F}_{1}^{\langle n;N \rangle} \delta_{m_{0}, \pm 1} \delta_{m_{1}, \mp 1},
\label{N_n_projection}
\eea
where the coefficients are given by
\bea
\mathcal{C}_{\jmath}^{\langle n;N \rangle}(\pm \lambda) &=& \exp \left( -\dfrac{\lambda ^{2}}{\Omega_{\jmath}^{2}}\right)
\dfrac{ 
	\left( \pm \dfrac{\lambda}{\Omega_{\jmath}} \right)^{n_{0}+ n_{1}+N_{0}+N_{1}} }
{2^{\frac{(N_{0}+N_{1})}{2}} \sqrt{n_{0}! n_{1}! N_{0}! N_{1}!}}, \quad \braket{n;N} \equiv (n_{0},n_{1};N_{0},N_{1}),\nn\\
\mathcal{F}_{0}^{\langle n;N \rangle} &=& \sum_{k=0}^{N_{0}} \sum_{\ell =0}^{N_{1}} (-1)^{\ell} 
\binom{N_{0}}{k} \binom{N_{1}}{\ell} 
{}_2F_0\Big(\! \! -n_{0},-k-\ell ;\phantom{}_{-} ; -\dfrac{\Omega_{0}^{2}}{\lambda ^{2}}\Big) \times \nn \\
&& \times\; {}_2F_0\Big(\! \! -n_{1},-N_{0}-N_{1}+k+\ell ;\phantom{}_{-} ; -\dfrac{\Omega_{0}^{2}}{\lambda ^{2}}\Big),\nn\\
\mathcal{F}_{1}^{\langle n;N \rangle} &=& \sum_{k=0}^{N_{0}} \sum_{\ell =0}^{N_{1}} (-1)^{k} 
\binom{N_{0}}{k} \binom{N_{1}}{\ell} 
{}_2F_0\Big(\! \! -n_{0},-k-\ell ;\phantom{}_{-} ; -\dfrac{\Omega_{1}^{2}}{\lambda ^{2}}\Big) \times \nn \\
&& \times\; {}_2F_0\Big(\! \! -n_{1},-N_{0}-N_{1}+k+\ell ;\phantom{}_{-} ; -\dfrac{\Omega_{1}^{2}}{\lambda ^{2}}\Big).
\label{C_F0_F1}
\eea
The hypergeometric sum in (\ref{C_F0_F1}) is defined as ${}_2F_0(\mathsf{x},\mathsf{y};\phantom{}_{-}; \tau) = \sum_{\ell=0}^{\infty} (\mathsf {x})_{\ell} (\mathsf{y})_{\ell}\, \frac{\tau^{\ell}}{\ell!}$, where the Pochhammer
symbol reads $(\mathsf {x})_{\ell} = \prod_{j = 0}^{\ell - 1}(\mathsf {x} + j)$. The projections (\ref{N_n_projection}) of the state 
$\ket{n_{0},m_{0};n_{1},m_{1}}$
facilitate its expansion  in the complete orthonormal basis set $\{\ket{\mathcal{E}_{\jmath}^{\braket{N}}}| \, \jmath \in (0, 1, \pm); N_{0}, N_{1} \in (0, 1, \ldots, \infty)\}$: 
\bea
\ket{n_{0},m_{0};n_{1},m_{1}} 
 =   \sum_{N_{0},N_{1} \atop =0}^{\infty} \; \sum_{\jmath \atop \in \{ 0,1,\pm \}}
 \mathfrak{C}_{\jmath }(\{ n,m;N\}) \,
  \ket{\mathcal{E}_{\jmath}^{\braket{N}}},
\label{n_E}  
\eea	
where the coefficients may be expressed as
\bea
\mathfrak{C}_{0}(\{ n,m;N \}) \!\!\!\! & \equiv & \!\!\!\! 
\braket{ \mathcal{E}_{0}^{\braket{N}} | n_{0},m_{0};n_{1},m_{1} } \nn \\
&=& \!\!\!\! \dfrac{1}{\sqrt{2}} \;
\Big( (-1)^{n_{0}+n_{1}+N_{1}} \delta_{m_{0},1} \delta_{m_{1},1} 
- (-1)^{N_{0}} \delta_{m_{0},-1} \delta_{m_{1},-1} \Big)\, 
\mathcal{C}_{0}^{\langle n;N \rangle}( \lambda) \; \mathcal{F}_{0}^{\langle n;N \rangle},\nn \\
\mathfrak{C}_{1}(\{ n,m;N \}) & \equiv & \braket{ \mathcal{E}_{1}^{\braket{N}} | n_{0},m_{0};n_{1},m_{1} } \nn \\
&=&  \!\!\!\! \dfrac{1}{\sqrt{2}} \;
\Big( (-1)^{n_{0}+N_{0}} \delta_{m_{0},1} \delta_{m_{1},-1} 
 - (-1)^{n_{1}+N_{1}} \delta_{m_{0},-1} \delta_{m_{1},1} \Big)\,
 \mathcal{C}_{1}^{\langle n;N \rangle}( \lambda) \; \mathcal{F}_{1}^{\langle n;N \rangle},\nn \\
\mathfrak{C}_{\pm}(\{ n,m;N \}) & \equiv & \braket{ \mathcal{E}_{\pm}^{\braket{N}} | n_{0},m_{0};n_{1},m_{1} } \nn \\
&=& \dfrac{1}{2} 
\sqrt{\dfrac{\chi_{_{\braket{N}}} \mp \Gamma_{-}}{\chi_{_{\braket{N}}}}} 
\Big( (-1)^{n_{0}+n_{1}+N_{1}} \delta_{m_{0},1} \delta_{m_{1},1}
 + (-1)^{N_{0}} \delta_{m_{0},-1} \delta_{m_{1},-1} \Big)  \times \nn \\
&& \times  \;
\mathcal{C}_{0}^{\langle n;N \rangle}( \lambda) \; \mathcal{F}_{0}^{\langle n;N \rangle} 
\pm \dfrac{1}{2}\dfrac{\Lambda _{\braket{N}}}{ \left| \Lambda _{\braket{N}} \right|} 
\sqrt{\dfrac{\chi_{_{\braket{N}}} \pm \Gamma_{-}}{\chi_{_{\braket{N}}}}} 
\Big( (-1)^{n_{0}+N_{0}} \delta_{m_{0},1} \delta_{m_{1},-1} \nn \\
 && + (-1)^{n_{1}+N_{1}} \delta_{m_{0},-1} \delta_{m_{1},1} \Big)\;
 \mathcal{C}_{1}^{\langle n;N \rangle}( \lambda) \; \mathcal{F}_{1}^{\langle n;N \rangle},
\label{C01+-} 
\eea	
The orthonormality of the state (\ref{n_E}) expressed in the basis set of the approximate energy eigenstates 
$\big\{\ket{\mathcal{E}_{\jmath}^{\braket{N}}}| \jmath \in \{ 0,1,\pm \},( N_{0},N_{1}) \in (0, 1, \ldots \infty)\big\}$ is 
confirmed via the hypergeometric identity
\bea
 \sum_{N_{0},N_{1}=0}^{\infty} \dfrac{1}{N_{0}! N_{1}!} S_{n_{0},n_{1}} (N_{0},N_{1}) S_{n'_{0},n'_{1}} (N_{0},N_{1}) 
 \left( \dfrac{\mathsf{x}}{2} \right)^{N_{0}+N_{1}}
=  \dfrac{n_{0}! n_{1}! }{\mathsf{x}^{n_{0} + n_{1}}} \exp(2 \mathsf{x}) \; \delta_{n_{0},n'_{0}} \; \delta_{n_{1},n'_{1}},
\label{hyper_identity}
\eea
where the bipartite weight functions read
\bea
S_{n_{0},n_{1}} (N_{0},N_{1}) &=& \sum_{k=0}^{N_{0}} \sum_{\ell=0}^{N_{1}} (-1)^{k} \binom{N_{0}}{k} \binom{N_{1}}{\ell}
{}_2F_0\Big(\! \! -n_{0},-k-\ell ;\phantom{}_{-} ; -\dfrac{1}{\mathsf{x}} \Big) \times \nn \\
&& \times \; {}_2F_0\Big(\! \! -n_{1},-N_{0}-N_{1}+k+\ell ;\phantom{}_{-} ; -\dfrac{1}{\mathsf{x}} \Big).
\label{weight_FF}
\eea
The expansion (\ref{n_E}) employing a complete set of energy eigenstates now readily yields the time evolution of the localized cavity states as follows:
\bea
\ket{n_{0},m_{0};n_{1},m_{1}}  \xrightarrow[\; \; t \; \;]{} \ket{\psi_{\braket{n,m}}(t)} 
 = \! \!  \sum_{N_{0},N_{1}  =0}^{\infty} \; \sum_{\jmath \atop \in \{ 0,1,\pm \}}
 \mathfrak{C}_{\jmath }(\{ n,m;N\})
 \exp \left( -i \mathcal{E}_{\jmath}^{\braket{N}} t \right) \,
  \ket{\mathcal{E}_{\jmath}^{\braket{N}}}.
\label{n0n1_t}  
\eea

\par

The setting specified above permits us now to explore the time evolution of a localized $N00N$-type of state residing in two cavities:
\beq
\ket{\psi(0)} = 
\dfrac{1}{\sqrt{1+|\mathrm{c}|^{2}}} \left( \ket{n,-1;0,-1} + \mathrm{c} \ket{0,-1;n,-1} \right), \qquad  
\mathrm{c} \in \mathbb{C}.
\label{N00N_0}
\eeq
The construction (\ref{n0n1_t}) immediately provides the subsequent transformation of the initial $N00N$-type of state (\ref{N00N_0}): 
\bea
\ket{\psi(t)} \! \! \! &=& \! \! \!  \dfrac{1}{\sqrt{1+|\mathrm{c}|^{2}}} \sum_{N_{0},N_{1}=0}^{\infty} (-1)^{N_{0}} \, \mathcal{C}_{0}^{( n,0;N_{0}, N_{1} )}(\lambda)
\left( \mathcal{F}_{0}^{( n,0;N_{0}, N_{1} )} + \mathrm{c} \,  \mathcal{F}_{0}^{ ( 0,n;N_{0}, N_{1} )} \right) \times \nn \\
&& \times \Bigg( \! \! \!   -\dfrac{1}{\sqrt{2}} \exp(-i\mathcal{E}_{0}^{\braket{N}}t )
\ket{\mathcal{E}_{0}^{\braket{N}}} + \dfrac{1}{2} 
\sqrt{\dfrac{\chi_{_{\braket{N}}} - \Gamma_{-}}{\chi_{_{\braket{N}}}}}  
 \exp(-i\mathcal{E}_{+}^{\braket{N}}t ) \ket{\mathcal{E}_{+}^{\braket{N}}} \nn \\
&&+ \dfrac{1}{2} 
\sqrt{\dfrac{\chi_{_{\braket{N}}} + \Gamma_{-}}{\chi_{_{\braket{N}}}}}  
 \exp(-i\mathcal{E}_{-}^{\braket{N}}t ) \ket{\mathcal{E}_{-}^{\braket{N}}} \Bigg). 
 \label{N00N_t}
\eea
The corresponding pure state density matrix is given by the usual tensorized prescription: $\rho(t) = \ket{\psi(t)}\bra{\psi(t)}$.

\section{The qubit reduced density matrix and its entanglement}
\label{reduced}
\setcounter{equation}{0}

In order to utilize the construction of the time-dependent state (\ref{N00N_t}) in studying the evolution of the entanglement between, say, the two qubits, we need to compose the relevant reduced density matrix using a partial tracing on the oscillator states:
\beq
\rho_{{}_{\mathcal{Q}}}(t) = \mathrm{Tr}_{{}_{\mathcal{O}}} \rho(t) \equiv 
\sum_{\imath , \jmath , k, \ell \atop \in \{ \pm 1 \}  }
\rho _{_{\imath, \jmath ; k, \ell}}(t) \ket{\imath \jmath} \bra{k \ell},
\label{red_den} 
\eeq
where the projection operators provide a complete basis set for the two qubit tensor product space. The elements 
$\rho _{_{\imath, \jmath ; k, \ell}}$ are conveniently expressed via a kernel whose structure follows from (\ref{N00N_t}):
\bea
G(\braket{N}; \braket{N'}) &=& \frac{1}{1+|\mathrm{c}|^{2}}\;  (-1)^{N_{0}+N'_{0}} \; 
\mathcal{C}_{0}^{( n,0;N_{0}, N_{1} )}(\lambda) \;\; \mathcal{C}_{0}^{(0, n;N'_{0}, N'_{1} )}(\lambda) \times \nn \\
&& \times \left( \mathcal{F}_{0}^{( n,0;N_{0}, N_{1} )} + \mathrm{c} \,  \mathcal{F}_{0}^{ ( 0,n;N_{0}, N_{1} )} \right)
 \left( \mathcal{F}_{0}^{( n,0;N'_{0}, N'_{1} )} + \mathrm{c}^{*} \,  \mathcal{F}_{0}^{ ( 0,n;N'_{0}, N'_{1} )} \right). 
 \label{G_N_Nprime}
\eea
It also obeys the appropriate Hermiticity and normalization properties:
\beq
 G(\braket{N^{\prime}};\braket{ N})^{*} = {G(\braket{N};\braket{ N^{\prime}})},\qquad
 \sum_{N_{0},N_{1} \atop =0}^{\infty} G(\braket{N};\braket{N}) = 1.
\label{herm_norm} 
\eeq

\par
The tools developed above now allow us to procure the time-dependent elements of the reduced density matrix of the qubits. We first enlist the real diagonal elements: 
\bea
\rho_{_{_{_{-1,-1;-1,-1}}}}(t) &=& \frac{3}{8} + \frac{1}{8} \sum_{N_{0},N_{1}=0}^{\infty} G(\braket{N}, \braket{N})
 \Bigg \lgroup \dfrac{1}{\chi ^{2}_{_{\braket{N}}}} 
\Bigg(  \Gamma ^{2}_{-} + 4 \Lambda ^{2}_{\braket{N}}
 \cos \left( \left(\mathcal{E}_{+}^{\braket{N}}-\mathcal{E}_{-}^{\braket{N}}\right) t \right) \Bigg)  \nn \\ 
 && + \dfrac{2}{\chi _{_{\braket{N}}}}  \Bigg( \left( \chi_{_{\braket{N}}} - \Gamma_{-} \right) 
 \cos \left( \left(\mathcal{E}_{0}^{\braket{N}}-\mathcal{E}_{+}^{\braket{N}}\right) t \right) \nn \\
 && + \left( \chi_{_{\braket{N}}} + \Gamma_{-} \right) 
 \cos \left( \left(\mathcal{E}_{0}^{\braket{N}}-\mathcal{E}_{-}^{\braket{N}}\right) t \right) \Bigg \rgroup, \nn \\
\rho_{_{_{_{-1,1;-1,1}}}}(t) &=& \dfrac{1}{2} \sum_{N_{0},N_{1}=0}^{\infty} G(\braket{N}, \braket{N}) 
\left( \dfrac{\Lambda _{\braket{N}}}{\chi_{_{\braket{N}}}} \right)^{2}
\Bigg ( 1 - \cos \left( \left(\mathcal{E}_{+}^{\braket{N}}-\mathcal{E}_{-}^{\braket{N}}\right) t \right) \Bigg), \nn \\
\rho_{_{_{_{1,-1;1,-1}}}}(t) &=& \dfrac{1}{2} \sum_{N_{0},N_{1}=0}^{\infty} G(\braket{N}, \braket{N}) 
\left( \dfrac{\Lambda _{\braket{N}}}{\chi_{_{\braket{N}}}} \right)^{2}
\Bigg ( 1 - \cos \left( \left(\mathcal{E}_{+}^{\braket{N}}-\mathcal{E}_{-}^{\braket{N}}\right) t \right) \Bigg), \nn \\
\rho_{_{_{_{1,1;1,1}}}}(t) &=& \dfrac{3}{8} + \dfrac{1}{8} \sum_{N_{0}N_{1}=0}^{\infty} G(\braket{N}, \braket{N})
\Bigg \lgroup \dfrac{1}{\chi ^{2}_{_{\braket{N}}}} 
\Bigg( \Gamma ^{2}_{-} + 4 \Lambda ^{2}_{\braket{N}}
\cos \left( \Big(\mathcal{E}_{+}^{\braket{N}}-\mathcal{E}_{-}^{\braket{N}}\Big) t \right) \Bigg) \nn \\ 
&& - \dfrac{2}{\chi _{_{\braket{N}}}} 
 \Bigg( \left( \chi_{_{\braket{N}}} - \Gamma_{-} \right) 
\cos \left( \left(\mathcal{E}_{0}^{\braket{N}}-\mathcal{E}_{+}^{\braket{N}}\right) t \right) \nn \\
&& + \left( \chi_{_{\braket{N}}} + \Gamma_{-} \right) 
\cos \left( \left(\mathcal{E}_{0}^{\braket{N}}-\mathcal{E}_{-}^{\braket{N}}\right) t \right) \Bigg \rgroup.
\label{den_1111}
\eea
The diagonal elements (\ref{den_1111}) ensure that trace of the qubit reduced density matrix is conserved: $\mathrm{Tr} \rho_{{}_{\mathcal{Q}}}(t) = 1$.
The evolution of the off-diagonal elements reflecting Hermiticity are entered below:
\bea
\!\!\!\! \!\!\!\! \rho_{_{_{_{-1,-1;-1,1}}}}(t) \!\!\!\! &=& \!\!\!\! \dfrac{1}{8} \sum_{N_{0},N_{1}=0}^{\infty} \sum_{N'_{0},N'_{1}=0}^{\infty}   
G(\braket{N};\braket{N'}) \left( \dfrac{\Lambda _{\braket{N'}}}{\chi_{_{\braket{N'}}}} \right)
\braket{N'_{0}|N_{0,-2}} \braket{N'_{1,-2}|N_{1}} \times \qquad \qquad \qquad \nn \\
& \times &  \!\!\!\! \!\! \Bigg \lgroup \dfrac{\chi_{_{\braket{N}}} - \Gamma_{-}}{\chi_{_{\braket{N}}}}
 \exp \left(-i\left(\mathcal{E}_{+}^{\braket{N}}-\mathcal{E}_{+}^{\braket{N'}}\right) t \right) -
 \dfrac{\chi_{_{\braket{N}}} + \Gamma_{-}}{\chi_{_{\braket{N}}}}
 \exp \left(-i\Big(\mathcal{E}_{-}^{\braket{N}}-\mathcal{E}_{-}^{\braket{N'}}\Big) t \right) \nn \\
& +&  \!\!\!\!  2 \exp \left(-i\Big(\mathcal{E}_{0}^{\braket{N}}-\mathcal{E}_{+}^{\braket{N'}}\Big) t \right)
- 2 \exp \left(-i(\mathcal{E}_{0}^{\braket{N}}-\mathcal{E}_{-}^{\braket{N'}}) t \right)
-  \dfrac{\chi_{_{\braket{N}}} - \Gamma_{-}}{\chi_{_{\braket{N}}}} \times \nn \\
&\times & \exp \left(-i\Big(\mathcal{E}_{+}^{\braket{N}}-\mathcal{E}_{-}^{\braket{N'}}\Big) t \right)  
+ \dfrac{\chi_{_{\braket{N}}} + \Gamma_{-}}{\chi_{_{\braket{N}}}}
\exp \left(-i\Big(\mathcal{E}_{-}^{\braket{N}}-\mathcal{E}_{+}^{\braket{N'}}\Big) t \right) \Bigg \rgroup, \nn
\eea
\bea
\!\!\!\! \!\!\!\! \!\!\!\! 
\rho_{_{_{_{-1,-1;1,-1}}}}(t) \!\!\!\! &=& \!\!\!\! \dfrac{1}{8} \sum_{N_{0},N_{1}=0}^{\infty} \sum_{N'_{0},N'_{1}=0}^{\infty}   
G(\braket{N};\braket{N'}) \left( \dfrac{\Lambda _{\braket{N'}}}{\chi_{_{\braket{N'}}}} \right)
\braket{N'_{0}|N_{0,-2}} \braket{N'_{1,2}|N_{1}} \times \nn \\
& \times &  \!\!\!\! \!\! \Bigg \lgroup \dfrac{\chi_{_{\braket{N}}} - \Gamma_{-}}{\chi_{_{\braket{N}}}}
\exp \left(-i\Big(\mathcal{E}_{+}^{\braket{N}}-\mathcal{E}_{+}^{\braket{N'}}\Big) t \right) -
\dfrac{\chi_{_{\braket{N}}} + \Gamma_{-}}{\chi_{_{\braket{N}}}}
\exp \left(-i\Big(\mathcal{E}_{-}^{\braket{N}}-\mathcal{E}_{-}^{\braket{N'}}\Big) t \right) \nn \\
& +&  \!\!\!\!  2 \exp \left(-i\Big(\mathcal{E}_{0}^{\braket{N}}-\mathcal{E}_{+}^{\braket{N'}}\Big) t \right)
- 2 \exp \left(-i\Big(\mathcal{E}_{0}^{\braket{N}}-\mathcal{E}_{-}^{\braket{N'}}\Big) t \right)
-  \dfrac{\chi_{_{\braket{N}}} - \Gamma_{-}}{\chi_{_{\braket{N}}}} \times \nn \\
& \times & \exp \left(-i\Big(\mathcal{E}_{+}^{\braket{N}}-\mathcal{E}_{-}^{\braket{N'}}\Big) t \right) 
+  \dfrac{\chi_{_{\braket{N}}} + \Gamma_{-}}{\chi_{_{\braket{N}}}}
\exp \left(-i\Big(\mathcal{E}_{-}^{\braket{N}}-\mathcal{E}_{+}^{\braket{N'}}\Big) t \right) \Bigg \rgroup, \nn
\eea
\bea
\!\!\!\! \!\!\!\! \rho_{_{_{_{-1,-1;1,1}}}}(t) \!\!\!\! &=& \!\!\!\! -\dfrac{1}{4} \sum_{N_{0},N_{1},N'_{0}=0}^{\infty} \!\!
G(N_{0},N_{1};N'_{0},N_{1})
\braket{N'_{0,2}|N_{0,-2}} 
\!\! \Bigg \lgroup \!\! \exp \left(-i \Omega_{0}(N_{0}-N'_{0}) t \right) \nn \\
 &+& \!\!\!\! \dfrac{\chi_{_{\braket{N}}} + \Gamma_{-}}{2 \chi_{_{\braket{N}}}} 
 \exp \left(-i\left(\mathcal{E}_{-}^{\braket{N}}-\mathcal{E}_{0}^{N'_{0},N_{1}}\right) t \right) 
 \! +  
 \dfrac{\chi_{_{\braket{N}}} - \Gamma_{-}}{2 \chi_{_{\braket{N}}}}
 \exp \left(-i\left(\mathcal{E}_{+}^{\braket{N}}-\mathcal{E}_{0}^{N'_{0},N_{1}}\right) t \right) \nn \\
 &-& \! \!\!\!\! \dfrac{\chi_{_{N'_{0},N_{1}}} \! + \! \Gamma_{-}}{2 \chi_{_{N'_{0},N_{1}}}} 
 \exp \! \left( \! -i \! \left( \! \mathcal{E}_{0}^{\braket{N}}-\mathcal{E}_{-}^{N'_{0},N_{1}}\right)\! t \! \right)
 \! \! - \! \dfrac{\chi_{_{N'_{0},N_{1}}} \! - \! \Gamma_{-}}{2 \chi_{_{N'_{0},N_{1}}}}
\exp \! \left( \! -i \! \left( \! \mathcal{E}_{0}^{\braket{N}}-\mathcal{E}_{+}^{N'_{0},N_{1}}\right)\! t \! \right) \nn \\
&-& \!\!\!\! \dfrac{\chi_{_{\braket{N}}} + \Gamma_{-}}{4 \chi_{_{\braket{N}}}} 
 \dfrac{\chi_{_{N'_{0},N_{1}}} + \Gamma_{-}}{\chi_{_{N'_{0},N_{1}}}}
 \exp \left(-i\left(\mathcal{E}_{-}^{\braket{N}}-\mathcal{E}_{-}^{N'_{0},N_{1}}\right) t \right) 
 -  \dfrac{\chi_{_{\braket{N}}} + \Gamma_{-}}{4 \chi_{_{\braket{N}}}} \;
   \dfrac{\chi_{_{N'_{0},N_{1}}} - \Gamma_{-}}{\chi_{_{N'_{0},N_{1}}}} \times \nn \\
& \times & \!\!\!\!   \exp \left(-i\Big(\mathcal{E}_{-}^{\braket{N}}-\mathcal{E}_{+}^{N'_{0},N_{1}}\Big) t \right) 
-  \dfrac{\chi_{_{\braket{N}}} - \Gamma_{-}}{4 \chi_{_{\braket{N}}}} \;
\dfrac{\chi_{_{N'_{0},N_{1}}} + \Gamma_{-}}{\chi_{_{N'_{0},N_{1}}}}
\exp \left(-i\Big(\mathcal{E}_{+}^{\braket{N}}-\mathcal{E}_{-}^{N'_{0},N_{1}}) t \right) \nn \\
&-& \!\!\!\!  \dfrac{\chi_{_{\braket{N}}} - \Gamma_{-}}{4 \chi_{_{\braket{N}}}} \;
\dfrac{\chi_{_{N'_{0},N_{1}}} - \Gamma_{-}}{\chi_{_{N'_{0},N_{1}}}}
\exp \left(-i\Big(\mathcal{E}_{+}^{\braket{N}}-\mathcal{E}_{+}^{N'_{0},N_{1}}\Big) t \right)
 \Bigg \rgroup, \nn
\eea
\bea
\rho_{_{_{_{-1,1;1,-1}}}}(t) \!\!\!\! &=& \!\!\!\! \dfrac{1}{4} \sum_{N_{0},N_{1},N'_{1}=0}^{\infty}
 \dfrac{\Lambda _{_{\braket{N}}} \Lambda _{_{N_{0},N'_{1}}}}{\chi_{_{\braket{N}}} \chi_{_{N_{0},N'_{1}}}}
 G(N_{0},N_{1};N_{0},N'_{1}) \braket{N'_{1,2}|N_{1,-2}} \times \nn \\
 & \times & \Bigg \lgroup \!\!
 \exp \left(-i\Big(\mathcal{E}_{+}^{\braket{N}}-\mathcal{E}_{+}^{N_{0},N'_{1}}\Big) t \right) 
+  \exp \left(-i\Big(\mathcal{E}_{-}^{\braket{N}}-\mathcal{E}_{+}^{N_{0},N'_{1}}\Big) t \right) \nn \\
&-&  \exp \left(-i\Big(\mathcal{E}_{+}^{\braket{N}}-\mathcal{E}_{-}^{N_{0},N'_{1}}\Big) t \right) 
-  \exp \left(-i\Big(\mathcal{E}_{-}^{\braket{N}}-\mathcal{E}_{+}^{N_{0},N'_{1}}\Big) t \right) \!\!\! \Bigg \rgroup, \quad 
\nn
\eea
\bea
\rho_{_{_{_{-1,1;1,1}}}}(t) \!\!\!\! &=& \!\!\!\! \dfrac{1}{8} \sum_{N_{0},N_{1}=0}^{\infty} \sum_{N'_{0},N'_{1}=0}^{\infty}   
G(\braket{N}; \braket{N'}) \left( \dfrac{\Lambda _{\braket{N}}}{\chi_{_{\braket{N}}}} \right)
\braket{N'_{0,2}|N_{0}} \braket{N'_{1}|N_{1,-2}} \times \nn \\
& \times &  \!\!\!\! \!\! \Bigg \lgroup \dfrac{\chi_{_{\braket{N'}}} - \Gamma_{-}}{\chi_{_{\braket{N'}}}}
\exp \left(-i\left(\mathcal{E}_{+}^{\braket{N}}-\mathcal{E}_{+}^{\braket{N'}}\right) t \right) -
\dfrac{\chi_{_{\braket{N'}}} + \Gamma_{-}}{\chi_{_{\braket{N'}}}}
\exp \left(-i\Big(\mathcal{E}_{-}^{\braket{N}}-\mathcal{E}_{-}^{\braket{N'}}\Big) t \right) \nn \\
& -&  \!\!\!\!  2 \exp \left(-i\Big(\mathcal{E}_{+}^{\braket{N}}-\mathcal{E}_{0}^{\braket{N'}}\Big) t \right)
+ 2 \exp \left(-i\Big(\mathcal{E}_{-}^{\braket{N}}-\mathcal{E}_{0}^{\braket{N'}}\Big) t \right)\nn \\
&+& \!\!\!\! \dfrac{\chi_{_{\braket{N'}}} + \Gamma_{-}}{\chi_{_{\braket{N'}}}}
\exp \left(-i\Big(\mathcal{E}_{+}^{\braket{N}}-\mathcal{E}_{-}^{\braket{N'}}\Big) t \right)  
-  \dfrac{\chi_{_{\braket{N'}}} - \Gamma_{-}}{\chi_{_{\braket{N'}}}}
\exp \left(-i\Big(\mathcal{E}_{-}^{\braket{N}}-\mathcal{E}_{+}^{\braket{N'}}\Big) t \right) \Bigg \rgroup, \nn
\eea
\bea
\!\!\!\!\! \!\! \rho_{_{_{_{1,-1;1,1}}}}(t) \!\!\!\! &=& \!\!\!\! \dfrac{1}{8} \sum_{N_{0},N_{1}=0}^{\infty} \sum_{N'_{0},N'_{1}=0}^{\infty}   
G(\braket{N};\braket{N'}) \left( \dfrac{\Lambda _{\braket{N}}}{\chi_{_{\braket{N}}}} \right)
\braket{N'_{0,2}|N_{0}} \braket{N'_{1}|N_{1,2}} \times \nn \\
& \times &  \!\!\!\! \!\! \Bigg \lgroup \dfrac{\chi_{_{\braket{N'}}} - \Gamma_{-}}{\chi_{_{\braket{N'}}}}
\exp \left(-i\Big(\mathcal{E}_{+}^{\braket{N}}-\mathcal{E}_{+}^{\braket{N'}}\Big) t \right) -
\dfrac{\chi_{_{\braket{N'}}} + \Gamma_{-}}{\chi_{_{\braket{N'}}}}
\exp \left(-i\Big(\mathcal{E}_{-}^{\braket{N}}-\mathcal{E}_{-}^{\braket{N'}}\Big) t \right) \nn \\
& -&  \!\!\!\!  2 \exp \left(-i\Big(\mathcal{E}_{+}^{\braket{N}}-\mathcal{E}_{0}^{\braket{N'}}\Big) t \right)
+ 2 \exp \left(-i\Big(\mathcal{E}_{-}^{\braket{N}}-\mathcal{E}_{0}^{\braket{N'}}\Big) t \right)\nn \\
&+ &\!\!\!\! \dfrac{\chi_{_{\braket{N'}}} + \Gamma_{-}}{\chi_{_{\braket{N'}}}}
\exp \left(-i\Big(\mathcal{E}_{+}^{\braket{N}}-\mathcal{E}_{-}^{\braket{N'}}\Big) t \right) 
-  \dfrac{\chi_{_{\braket{N'}}} - \Gamma_{-}}{\chi_{_{\braket{N'}}}}
\exp \left(-i\Big(\mathcal{E}_{-}^{\braket{N}}-\mathcal{E}_{+}^{\braket{N'}}\Big) t \right) \Bigg \rgroup \! .
\label{off_diagonal}
\eea
Our evaluation (\ref{off_diagonal}) of the off-diagonal elements of the reduced density matrix employs  the
following scalar products of the shifted number states of the delocalized oscillators: 
\bea
\braket{M_{\jmath,-2}|N_{\jmath,2}} &=&  (-1)^{N+M} \braket{M_{\jmath,2}|N_{\jmath,-2}} \nn\\
&= &\dfrac{(-1)^{M}}{\sqrt{M! N!}} 
\left( \dfrac{2 \sqrt{2} \lambda}{\Omega_{\jmath}} \right)^{M+N} \exp \left(-\dfrac{4 \lambda ^{2}}{\Omega_{\jmath} ^{2}} \right) 
{}_2F_0\Big(\! \! -M,-N ;\phantom{}_{-} ; -\dfrac{\Omega_{\jmath}^{2}}{8 \lambda ^{2}}\Big), \nn\\
\braket{M_{\jmath}|N_{\jmath,2}} &=& \braket{M_{\jmath,-2}|N_{\jmath}} = (-1)^{N+M} \braket{M_{\jmath,2}|N_{\jmath}} = 
(-1)^{N+M}\braket{M_{\jmath}|N_{\jmath,-2}}\nn\\
&=&\dfrac{(-1)^{M}}{\sqrt{M! N!}} 
\left( \dfrac{\sqrt{2} \lambda}{\Omega_{\jmath}} \right)^{M+N} \exp \left(-\dfrac{ \lambda ^{2}}{\Omega_{\jmath} ^{2}} \right) 
{}_2F_0\Big(\! \! -M,-N ;\phantom{}_{-} ; -\dfrac{\Omega_{\jmath}^{2}}{2 \lambda ^{2}}\Big). 
\label{scaler_prod}
\eea
\par

\begin{figure}[H]
	\captionsetup[subfigure]{labelformat=empty}
	\captionsetup[subfigure]{labelformat=empty} 
	\subfloat[(\sf{a})]{\includegraphics[width=5.3cm,height=3.5cm]
		{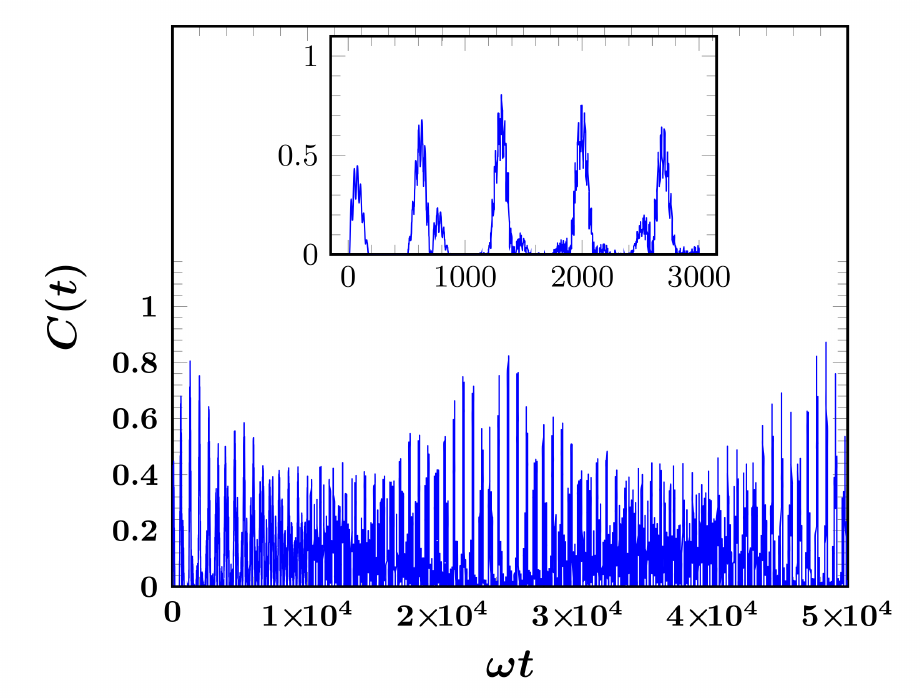}}
	\captionsetup[subfigure]{labelformat=empty}
	\subfloat[(\sf{b})]{\includegraphics[width=5.3cm,height=3.5cm]
		{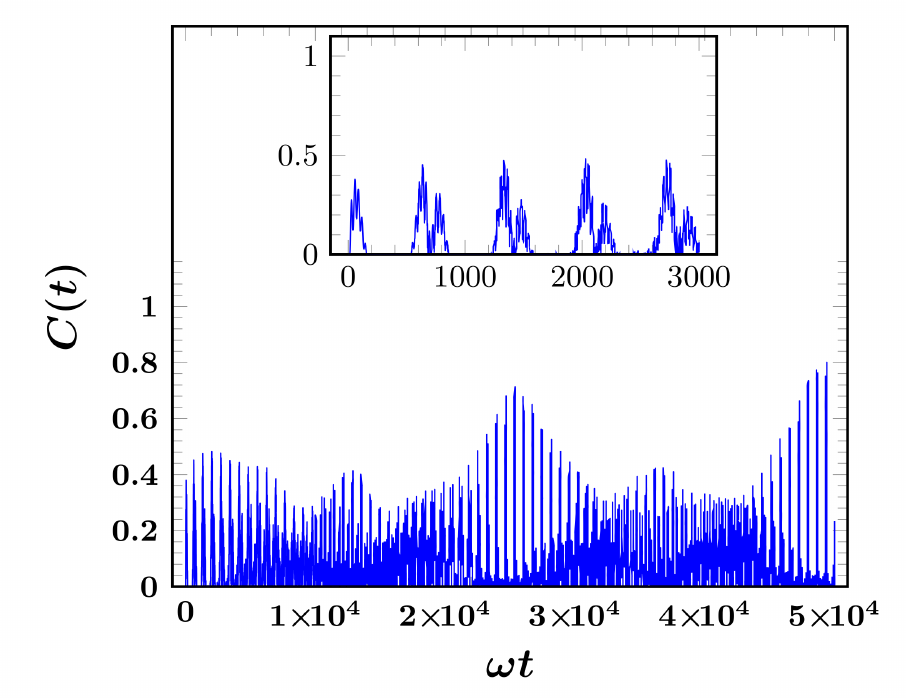}} 
	\captionsetup[subfigure]{labelformat=empty}
	\subfloat[(\sf{c})]{\includegraphics[width=5.3cm,height=3.5cm]
		{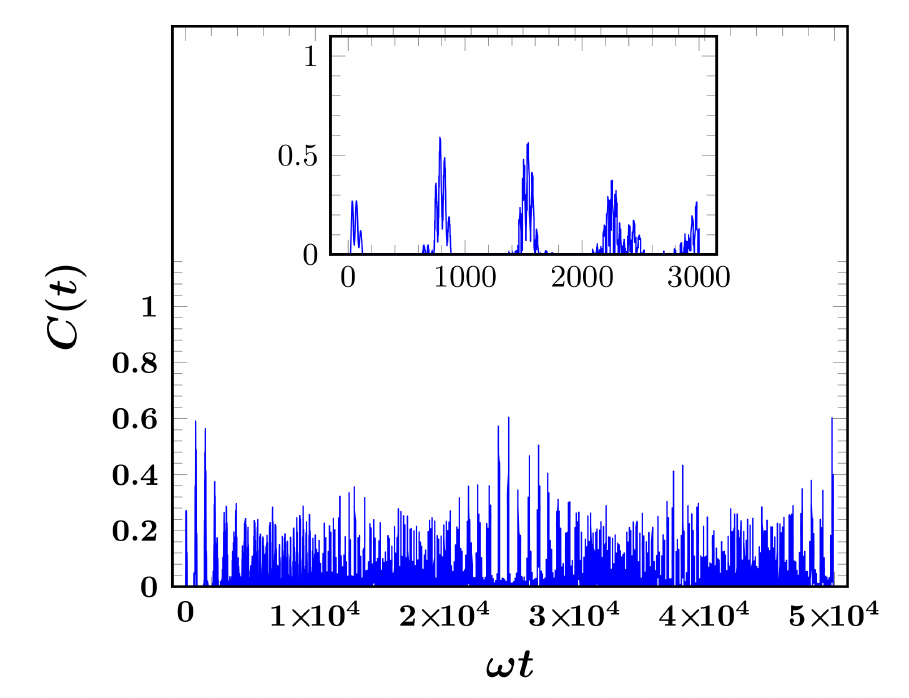}}
\caption{For the parametric choices $ \mathrm{c}=i, \omega=1, \lambda=0.1, \Delta=0.15, \nu=0.5$,  
the figures ({\sf a, b, c}) refer to  the ascending eigenvalues of the localized  number operator $n=4,6,10$, respectively. Sudden disappearance of the entanglement between the qubits, while observed for all values of $n$, becomes more prolonged with the
increasing photon number.}
\label{concur_evolve}
\end{figure}

With the explicit description of the reduced density matrix of the two-qubit system in hand we now turn towards studying its
entanglement properties. To determine the extent of entanglement between the qubits  we use the concurrence which is  widely accepted as its  measure for the bipartite mixed states. The concurrence introduced by Wootters [\cite{W1998}] is defined as
\beq
 C(t) = \max \big\{0,\sqrt{\lambda_{1}}-\sqrt{\lambda_{2}}-\sqrt{\lambda_{3}}-\sqrt{\lambda_{4}}\big\},
 \label{concurrence}
\eeq
where $\big\{\lambda_{\imath}|\, \imath = (1, \ldots, 4)\big\}$ are the eigenvalues, ordered in the descending sequence, of the matrix 
\beq
{\mathsf R}(t) = \rho _{{}_{\mathcal{Q}}}(t) \widetilde{\varrho}(t), \qquad
\widetilde{\varrho}(t) = \left( \sigma^{\mathrm {y}}  \otimes \sigma^{\mathrm{y}} \right) \rho ^{*}_{{}_{\mathcal{Q}}}(t)  \left( \sigma^{\mathrm {y}}  \otimes \sigma^{\mathrm{y}} \right).
\label{R_t}
\eeq
The matrix  $\widetilde{\varrho}(t)$ results from the spin-flip operation on the reduced qubit density matrix 
$\rho_{{}_{\mathcal{Q}}}(t)$. The two-qubit system remains entangled for $C(t) > 0$. The maximum possible entanglement is
achieved at the limiting value $C(t) = 1$, while $ C(t) = 0$ implies separability. We now examine the evolution of the entanglement of the bipartite reduced density matrix (\ref{red_den}) as quantified by the measure (\ref{concurrence}). 
We notice that for the higher values of the localized photon numbers ($n$) of the $N00N$-type states (\ref{N00N_0}) the 
variation of the concurrence of the spin degrees of freedom with the scaled time (Fig. \ref{concur_evolve} {\sf a, b, c}) depicts the sudden death [\cite{YE2006}, \cite{YYE2006}], and the disappearance of the two-qubit entanglement for a comparatively longer period. The qubit-oscillator interaction
ensures that informations carried by the phase correlation between the qubits passes away to the oscillator degrees of freedom.
The larger the size of the $N00N$-type of the oscillator state, the information will take a more prolonged time to reappear 
in the qubit subsystem and rejuvenate the entanglement between the two qubits. 

\par

\begin{figure}[H]
	\captionsetup[subfigure]{labelformat=empty} 
	\subfloat[(\sf a)]{\includegraphics[width=8.3cm,height=3.5cm]
		{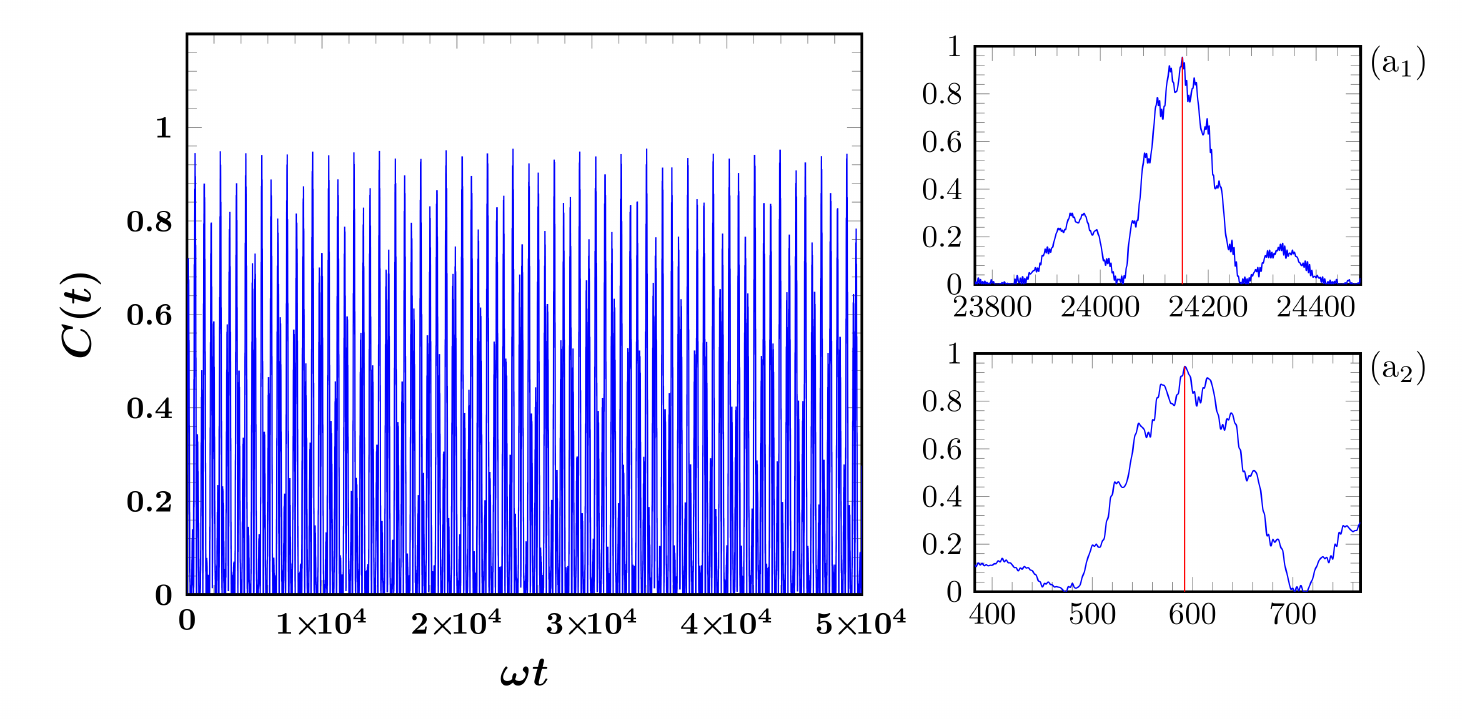}}
\captionsetup[subfigure]{labelformat=empty}
	\subfloat[({\sf b})]{\includegraphics[width=8.3cm,height=3.5cm]
	{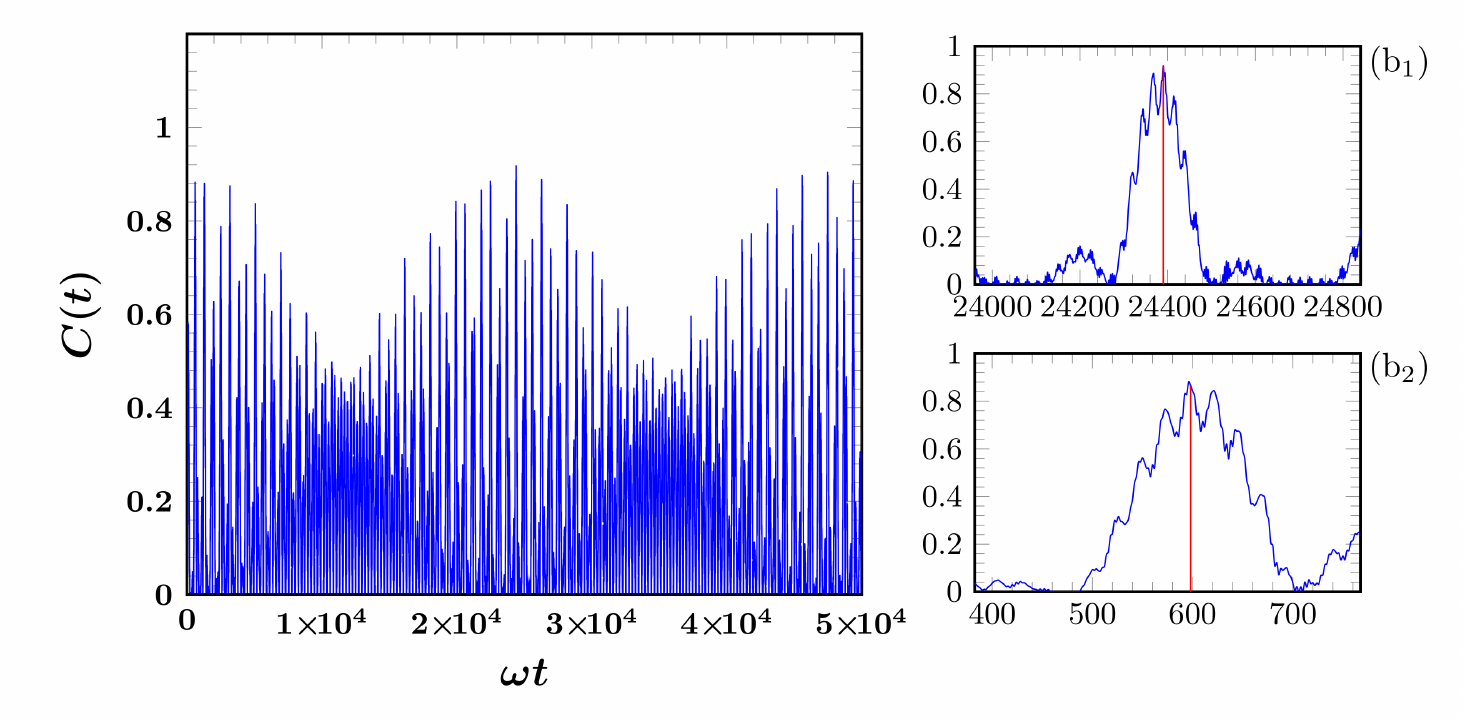}}
   \caption{For the present set of diagrams we make the following parametric choices:  $\mathrm{c}=0, \omega=1, \lambda=0.1, \Delta=0.15, \nu=0.5$. The graphs (${\mathsf{a}, \mathsf a_{1}},{\mathsf a_{2}}$) plot the concurrence $C(t)$ 
     for the value $n = 1$ in the initial state (\ref{N00N_0}), whereas the illustrations 
   (${\mathsf{b}, \mathsf b_{1}},{\mathsf b_{2}}$) involve the corresponding selection $n = 2$. The values of the concurrence $C(t)$ at  its local maxima, and the pertinent construction of the nearly pure state two-qubit density matrices $ \ket{\Phi}\bra{\Phi}$
   described earlier are reported in Table \ref{tab}.} 
\label{high_concurrence}
\end{figure}

Lastly, we use our evaluation of the concurrence $C(t)$ (\ref{concurrence}) to produce two-qubit states which are in the close neighborhood of the maximally entangled  generalized Bell states. Towards this, we consider (Figs. \ref{high_concurrence} 
({\sf a, b})) the qubit reduced density matrix $\rho _{{}_{\mathcal{Q}}}(t)$ at instants corresponding to the dominant values of the concurrence $C(t) \lesssim 1$, and minimize its Hilbert-Schmidt distance [\cite{DMMW2000}] from a pure state density matrix 
$\ket{\Phi}\bra{\Phi}$:
\beq
\mathrm{d}_{\mbox{\tiny{HS}}} = \sqrt{\mathrm{Tr}\,(\rho_{_{\mathcal{Q}}} - \rho_{_{\ket{\Phi}}})^{2}}, \;
\rho_{_{\ket{\Phi}}} = \ket{\Phi}\bra{\Phi}, \; \ket{\Phi} = \alpha \ket{\phi_{+}} + \beta \ket{\phi_{-}} + \gamma \ket{\varphi_{+}} + \delta \ket{\varphi_{-}},
\label{hilbert}
\eeq
where the generalized Bell basis states read
\beq
\ket{\phi_{\pm}} = \dfrac{1}{\sqrt{2}} (\ket{1,1} \pm i \ket{-1,-1}), \quad
\ket{\varphi_{\pm}} = \dfrac{1}{\sqrt{2}} (\ket{1,-1} \pm i \ket{-1,1}).
\label{gen_bell}
\eeq
The above coefficients $((\alpha, \beta, \gamma, \delta) \in \mathbb{C})$  maintaining the normalization 
$(|\alpha|^{2} + |\beta|^{2} + |\gamma|^{2} +|\delta|^{2} =1)$ are varied to detect a linear combination 
 of the generalized Bell states (\ref{gen_bell}) that minimizes the distance 
(\ref{hilbert}). To emphasize the dynamical effects that produce the entangled two-qubit almost pure states, we, in this
instance (Fig. \ref{high_concurrence}), adopt the choice  $\mathrm{c}=0$ in the initial state (\ref{N00N_0}) imparting an unentangled factorized structure to it.

\par

As high concurrence (\ref{concurrence}) limits are evinced more frequently for low-lying photon number states, we set the values $n = 1$ and $n = 2$ for the initial state (\ref{N00N_0}) in the description of the Figs. \ref{high_concurrence} ({\sf a}) and ({\sf b}), consecutively. Diagrams (${\mathsf a_{1}},{\mathsf a_{2}}$) and
(${\mathsf b_{1}},{\mathsf b_{2}}$) specify the  time slices in Figs. \ref{high_concurrence} ({\sf a}) and ({\sf b}), respectively, at which the local peaks in the concurrence $C(t)$ are studied. The Hilbert-Schmidt distance $\mathrm{d}_{\mbox{\tiny{HS}}}$ (\ref{hilbert}) is minimized over the ensemble of states $\{\ket{\Phi}| (\alpha, \beta, \gamma, \delta) \in \mathbb{C}\}$ obtained via the variations in the said complex coefficients. The relevant quantities and the characterization of the states engendering minimum distance $\mathrm{d}_{\mbox{\tiny{HS}}}$ are registered in Table \ref{tab}. For  the parametric range considered here we notice that  at the instants, when the local maxima of the concurrence $C(t)$ are realized, the resultant qubit reduced density matrices are predominantly majorized by the pure generalized Bell state density matrix 
$\rho_{_{\mathcal{Q}}} \sim  \ket{\phi_{\pm}}  \bra{\phi_{\pm}}$. It is interesting to note that 
in the  two-qubit states observed in Figs. \ref{high_concurrence} ({\sf a, b}), the relative phases equaling $\pm \frac{\pi}{2}$ appear between the components $\ket{1,1}$ and $\ket{-1,-1}$ signifying the introduction of an effective magnetic field.
\begin{table}[h]
	\centering
	\begin{tabular}{c|M{1.1in}|M{1.1in}|M{1.1in}|M{1.1in}|}
		\cline{2-5}
		& \multicolumn{2}{c|}{$n=1$} & \multicolumn{2}{c|}{$n=2$} \\ \hline
		\multicolumn{1}{|M{1.2in}|}{$\omega t$} &    $592$         &     $24152$       &      $598$       &   $24390$         \\ \hline
		\multicolumn{1}{|M{1.2in}|}{$C(t)$}    &      $0.945267$       &       $0.954379$     &     $0.865998$        &   $0.917634$         \\ \hline
		\multicolumn{1}{|M{1.2in}|}{$\mathrm{d}_{\mbox{\tiny{HS}}} \lvert_{\min} $}    &   $0.089865$          &    $0.029438$        &   $0.110694$          &      $0.0836728$      \\ \hline
		\multicolumn{1}{|M{1.2in}|}{$ \ket{\Phi} $}  & \parbox{1.2in}{\centering $0.997136 \ket{\phi_{+}} - 0.0292865 \ket{\phi_{-}} + 0.0697298 \ket{\varphi_{-}} $ \\  $ \approx \ket{\phi_{+}}  $ } &    \parbox{1.2in}{\centering $0.0099995 \ket{\phi_{+}} +0.999950 \ket{\phi_{-}} $ \\  $ \approx \ket{\phi_{-}}$ }       & \parbox{1.2in}{\centering $0.996483 \ket{\phi_{+}}  - 0.0560522 \ket{\phi_{-}} - 0.0622802 \ket{\varphi_{-}} $ \\ $\approx \ket{\phi_{+}}$  }      & \parbox{1.2in}{\centering $0.999236 \ket{\phi_{-}}  - 0.0390937 \ket{\varphi_{+}}$ \\ $\approx \ket{\phi_{-}}$}      \\ \hline
	\end{tabular}
	\caption{}
	\label{tab}
\end{table}
\section{Conclusion}
\label{conclu}
We considered a two cavity system where each cavity included a qubit and an oscillator degrees of freedom strongly interacting with each other. The tunneling  of photons between the cavities is permitted. The dominant oscillator frequency controls the
slow moving qubit allowing us to use an adiabatic approximation. Setting up the delocalized variables for the oscillators and the qubits  we approximately diagonalize the Hamiltonian. Starting with a $N00N$-type initial state we examine its evolution and construct, for instance, the reduced density matrix of the bipartite system of two qubits. Utilizing the concurrence  the entanglement of the two-qubit reduced density is measured. For the high $N00N$-type initial states the phenomenon of sudden death and the continued absence of entanglement between the qubits are increasingly visible as the macroscopic size of the entangled photonic states prevents the information passing to the qubits for a longer time. On the other hand for a low value
of the photon quantum number of the initial state nearly maximal entanglement is reached, at certain times, for the two-qubit states that behave as almost pure generalized Bell states. Our analysis of the time-development of the initial state makes it possible to extract 
the reduced density matrix for the two oscillator degrees of freedom, and thereby study the multidimensional phase space quasiprobability distributions. This  will be pursued elsewhere.      
\section*{Acknowledgements}
One of us (RC) wishes to thank the
Quantum Optics and Quantum Information Theory group in the Institute of Mathematical Sciences and the Department of Nuclear Physics, University of Madras for kind hospitality. Another author (VY) acknowledges the support from Department of Science 
and Technology (India) under the INSPIRE Fellowship scheme. We are also indebted for generous computational help from the Department of Central Instrumentation and Service Laboratory, and the Department of Nuclear Physics, University of Madras.
We are happy to express our sincere gratitude to B. Virgin Jenisha for numerous discussions.  


\begin{thebibliography}{99}
\bibitem{AKSV2003}  D.K. Armani, T.J. Kippenberg, S.M. Spillane, K.J. Vahala, Nature {\bf 421} (2003) 925.

\bibitem{BTO2000} M. Bayindir, B. Temelkuran, E. Ozbay, Phys. Rev. Lett. {\bf 84} (2000) 2140.

\bibitem{ABS2002} A.D. Armour, M.P. Blencowe, K.C. Schwab, Phys. Rev. Lett. {\bf 88} (2002) 148301.

\bibitem{WALLRAFF2004} A. Wallraff, D.I. Schuster, A. Blais, L. Frunzio, R.S. Huang,
J. Majer, S. Kumar, S.M. Girvin, R.J. Schoelkopf, Nature {\bf 431} (2004), 162.

\bibitem{CEHM1999}  J.I. Cirac, A.K. Ekert, S.F. Huelga, C. Macchiavello, Phys. Rev. A {\bf 59} (1999) 4249.

\bibitem{AB2007} D.G. Angelakis, S. Bose, J. Opt. Soc. Am. B {\bf 24} (2007) 266.

\bibitem{BAB2007} S. Bose, D.G. Angelakis, D. Burgarth, J. Mod. Opt. {\bf 54} (2007) 2307.

\bibitem{OIK2008} C.D. Ogden, E.K. Irish, M.S. Kim, Phys. Rev. A {\bf 78} (2008) 063805.

\bibitem{CS2011} R. Chakrabarti, G. Sreekumari, J. Phys. B {\bf 44} (2011) 115505.

\bibitem{HBP2007}  M.J. Hartmann, F.G.S.L. Brand$\tilde{\mathrm{a}}$o, M.B. Plenio, Phys. Rev. Lett. {\bf 99} (2007) 160501.

\bibitem{AK2008} D.G. Angelakis, A. Kay, New J. Phys. {\bf 10} (2008) 023012.

\bibitem{JC1963} E.T. Jaynes, F.W. Cummings, Proc. IEEE {\bf 51} (1963) 89.

\bibitem{TODOROV2009} Y. Todorov, A.M. Andrews, I. Sagnes, R. Colombelli, P. Klang, G. Strasser, C. Sirtori, Phys. Rev. Lett.
{\bf 102} (2009) 186402.

\bibitem{HIF2012} S. Hayashi, Y. Ishigaki, M. Fujii, Phys. Rev. B {\bf 86} (2012) 045408.

\bibitem{GUNTER2009} G. G\"{u}nter, A.A. Anappara, J. Hees, A. Sell, G. Biasiol, L. Sorba, S. De Liberato, C. Ciuti, A. Tredicucci, A. Leitenstorfer, R. Huber, Nature {\bf 458}, 178 (2009).

\bibitem {AAA2009}  A.A. Anappara, S. De Liberato, A. Tredicucci, C. Ciuti, G. Biasiol, L. Sorba, F. Beltram, 
Phys. Rev. B {\bf 79} (2009) 201303 (R).

\bibitem{LSESR2009}  M.D. LaHaye, J. Suh, P.M. Echternach, K.C. Schwab, M.L. Roukes, Nature Lett. {\bf 459} (2009) 960.

\bibitem{HOFHEINZ2009} M. Hofheinz, H. Wang, M. Ansmann, R.C. Bialczak, E. Lucero, M. Neeley, A.D. O'Connell, D. Sank,
J. Wenner, J.M. Martinis, A.N. Cleland, Nature {\bf 459} (2009) 546.

\bibitem{NIEMCZYK2010} T. Niemczyk, F. Deppe, H. Huebl, E.P. Menzel, F. Hocke, M.J. Schwarz, J.J. Garcia-Ripoll, D. Zueco, T. H\"{u}mmer, E. Solano, A. Marx, R. Gross, Nat. Phys. {\bf 6} (2010) 772.

\bibitem{YN2005}J.Q. You, F. Nori, Phys. Today {\bf 58} (2005) 42.

\bibitem{BAN2011} I. Buluta, S. Ashhab, F. Nori, Rep. Prog. Phys. {\bf 74} (2011) 104401.

\bibitem{GAN2014} I.M. Georgescu, S. Ashhab, F. Nori, Rev. Mod. Phys. {\bf 86} (2014) 153.

\bibitem{ZAYN2013} Z.L. Xiang, S. Ashhab, J.Q. You, F. Nori, Rev. Mod. Phys. {\bf 85} (2013) 623.

\bibitem{IGMS2005} E.K. Irish, J. Gea-Banacloche, I. Martin, K.C. Schwab, Phys. Rev. B {\bf 72} (2005) 195410.

\bibitem{AN2010} S. Ashhab, F. Nori, Phys. Rev. A {\bf 81} (2010) 042311.

\bibitem{YZZ2012} P. Yang, P. Zou, Z.-M. Zhang, Phys. Lett. A {\bf 376} (2012) 2977.

\bibitem{D2016}  K. Dong, Chin. Phys. B. {\bf 25} (2016) 124202.

\bibitem{SCWY2014} L.-T. Shen, R.-X. Chen, H.-Z. Wu, Z.-B. Yang, Phys. Rev. A {\bf 89} (2014) 023810. 

\bibitem{BIWH1996} J.J. Bollinger, W.M. Itano, D.J. Wineland, D.J. Heinzen, Phys. Rev. A {\bf 54} (1996) R4649. 

\bibitem{KLD2002} P. Kok, H. Lee, J.P. Dowling, Phys. Rev. A {\bf 65} (2002) 052104. 

\bibitem{MLS2004} M.W. Mitchell, J.S. Lundeen, A.M. Steinberg, Nature {\bf 429} (2004) 161.

\bibitem{BOTO2000} A.N. Boto, P. Kok, D.S. Abrams, S.L. Braunstein, C.P. Williams, J.P. Dowling, Phys. Rev. Lett. 
{\bf 85} (2000) 2733.

\bibitem{NOOST2007} T. Nagata, R. Okamoto, J.L. O'Brien, K. Sasaki, S. Takeuchi, Science {\bf 316} (2007) 726.

\bibitem{AAS2011} I. Afek, O. Ambar, Y. Silberberg, Science {\bf 328} (2010) 879.

\bibitem{Wang2011} H. Wang, M. Mariantoni, R.C. Bialczak, M. Lenander, E. Lucero, M. Neeley, A.D. O'Connell, 
D. Sank, M. Weides, J. Wenner, T. Yamamoto, Y. Yin, J. Zhao, J.M. Martinis, A.N. Cleland, 
Phys. Rev. Lett. {\bf 106} (2011) 060401.

\bibitem{W1998}  W.K. Wootters, Phys. Rev. Lett. {\bf 80} (1998) 2245.

\bibitem{YE2006} T. Yu, J.H. Eberly, Phys. Rev. Lett. {\bf 97} (2006) 140403.

\bibitem{YYE2006} M. Y\"{o}na\c{c}, T. Yu, J.H. Eberly, J. Phys. B {\bf 39} (2006) S621.

\bibitem{DMMW2000} V.V. Dodonov, O.V. Man'ko, V.I. Man'ko, A. W\"{u}nsche, J. Mod. Opt. {\bf 47} (2000) 633.
\end{thebibliography}
\end{document}